\documentclass[sigconf,screen,authorversion]{acmart}

\usepackage{soul}
\usepackage{natbib}
\usepackage{CJKutf8}
\usepackage{multicol,caption}

\AtBeginDocument{%
  }

\copyrightyear{2025}
\acmYear{2025}
\setcopyright{rightsretained}
\acmConference[CHI EA '25]{Extended Abstracts of the CHI Conference on
Human Factors in Computing Systems}{April 26-May 1, 2025}{Yokohama, Japan}
\acmBooktitle{Extended Abstracts of the CHI Conference on Human Factors in
Computing Systems (CHI EA '25), April 26-May 1, 2025, Yokohama,
Japan}\acmDOI{10.1145/3706599.3720135}
\acmISBN{979-8-4007-1395-8/2025/04}

\begin{document}
\begin{CJK*}{UTF8}{ipxm}
\title[More-than-Human Storytelling]{More-than-Human Storytelling: Designing Longitudinal Narrative Engagements with Generative AI}

\author{\'Emilie Fabre}
\email{fabre@g.ecc.u-tokyo.ac.jp}
\orcid{0000-0003-1978-9374}
\affiliation{%
  \institution{The University of Tokyo}
  \city{Tokyo}
  \country{Japan}
}

\author{Katie Seaborn}
\email{seaborn.k.aa@m.titech.ac.jp}
\orcid{0000-0002-7812-9096}
\affiliation{%
  \institution{Institute of Science Tokyo}
  \city{Tokyo}
  \country{Japan}
}

\author{Shuta Koiwai}
\email{koiwai.s.aa@m.titech.ac.jp}
\orcid{0009-0007-2457-5983}
\affiliation{%
  \institution{Institute of Science Tokyo}
  \city{Tokyo}
  \country{Japan}
}

\author{Mizuki Watanabe}
\orcid{0000-0003-3036-7041}
\affiliation{%
  \institution{Institute of Science Tokyo}
  \city{Tokyo}
  \country{Japan}
}
\email{watanabe.m.ca@m.titech.ac.jp}

\author{Paul Riesch}
\email{riesch.p.research@gmail.com}
\orcid{0009-0003-3666-1564}
\affiliation{%
  \institution{Institute of Science Tokyo}
  \city{Tokyo}
  \country{Japan}
}
\affiliation{%
  \institution{Stuttgart Media University}
  \city{Stuttgart}
  \country{Germany}
}

\renewcommand{\shortauthors}{Fabre et al.}

\begin{abstract}
Longitudinal engagement with generative AI (GenAI) storytelling agents is a timely but less charted domain. We explored multi-generational experiences with ``Dreamsmithy,'' a daily dream-crafting app, where participants ($N=28$) co-created stories with AI narrator ``Makoto'' every day. Reflections and interactions were captured through a two-week diary study. Reflexive thematic analysis revealed themes likes ``oscillating ambivalence'' and ``socio-chronological bonding,'' highlighting the complex dynamics that emerged between individuals and the AI narrator over time. Findings suggest that while people appreciated the personal notes, opportunities for reflection, and AI creativity, limitations in narrative coherence and control occasionally caused frustration. The results underscore the potential of GenAI for longitudinal storytelling, but also raise critical questions about user agency and ethics. We contribute initial empirical insights and design considerations for developing adaptive, more-than-human storytelling systems.
\end{abstract}




\maketitle

\section{Introduction}
New and improved artificial intelligence (AI) models are arriving every month, each delivering ever-increasing functionality, ease of use, and speed~\cite{Kumar2023}. In the midst of all this progress, human-computer interaction (HCI) researchers are still playing catch-up, analyzing how and why people are using these new tools~\cite{Sauvola2024,Venkatesh2021}.
AI, and specifically Generative AI (GenAI), has its share of criticism and ethical concerns. Still, the technology has been taken up wide-scale within public and professional spaces, and seems to be here to stay. Many successful products and open source projects are already available to the general public, allowing those even without advanced technical skills to generate text, speech, music, images, and videos, often in seconds. Of these, \emph{text generation} leads use of GenAI. Notable is Open AI's flagship ChatGPT large language model (LLM), one of the most popular tools available globally, which boasts more than 100 millions weekly users as of 2023\footnote{\url{https://www.reuters.com/technology/chatgpt-sets-record-fastest-growing-user-base-analyst-note-2023-02-01/}}. ChatGPT and other text-based GenAI tools are becoming commonplace in education~\cite{Jo2024}, software engineering~\cite{chatgptsoftware}, and daily life activities~\cite{WOLF2024102821}, including for news, translation, and ideation.

Among these use cases, one of the most controversial is the place of GenAI in \emph{artistic fields}~\cite{Epstein_2023art} and notably \emph{storytelling and script-writing}~\cite{Han_2023story}. The film industry in particular saw strike movements from various guilds and unions pushing for clear rules around GenAI following increasing interest in its use from production companies\footnote{\url{https://www.theguardian.com/culture/2023/oct/01/hollywood-writers-strike-artificial-intelligence}}. Many concerns also exist in terms of the ethics of the data that are used to train LLMs, especially with companies using data without appropriate permission\footnote{\url{https://www.theverge.com/2025/1/14/24343692/meta-lawsuit-copyright-lawsuit-llama-libgen}}.
Nevertheless, several products making use of text GenAI are already on the market. Examples include SillyTavern\footnote{\url{https://sillytavernai.com/}}, NovelAI\footnote{\url{https://novelai.net/}}, RisuAI\footnote{\url{https://risuai.net/}}, c.ai\footnote{\url{https://character.ai/}}, and many more. The main draw of these tools lie in the user being able to tweak and customize their experience, creating new or continuing ever changing stories with each use. As yet, the effects of repeated exposure to such stories, whether and how users engage with these tools over time, and how 
users can be involved in the GenAI's creative process is not well understood~\cite{Han_2023story,Chu_2024story}.


As a first step, we explored a novel multi-day GenAI-based storytelling engagement. We used a diary study~\cite{fan2024diary,Bolger_2003diary} to elicit timely and contextual accounts of older and younger people's experiences with the agent at home. We aimed to provide rich accounts of and insights on longitudinal experiences with storytelling GenAI. We asked: \textbf{What user experiences can GenAI-based longitudinal storytelling agents provide?}
We offer these contributions:

\begin{itemize}
    \item \textbf{Research}: We provide empirical insights into how users engage with GenAI storytelling agents longitudinally. Our study identifies eight key themes, such as oscillating ambivalence, uncanny creativity, and socio-chronological bonding, that shape user experiences with AI-driven narrative interactions over time.
    \item \textbf{Methodological}: We used Reflexive Thematic Analysis \cite{Braun2023} on a daily, 14 day experience with an AI story agent.
    \item \textbf{Practical}: We offer design considerations for developing personalized, engaging, and adaptive AI storytelling agents.
\end{itemize}

\section{Background}
\subsection{GenAI in Storytelling}
Even before the rise of LLMs, researchers noticed the interest by some in artistic fields to use AI as a creative writing tool. In 2020, \citet{heysiristory} presents an excerpt of a story generated by an early Transformer model. The author analyzed it and other uses of early AI storytelling experiences. They concluded that the warped and uncanny perspectives and narratives coming from the AI could either become the seed of new, creative, unbiased ways for users to reflect upon the world or yet another way for companies and commercial apps to create ultra-tailored content akin to digital marketing.
\citet{heysiristory} is also notable, being published a month before the release of GPT-3, the model that would later become the basis for ChatGPT. Here, both the optimism that emanated from this early ``Wild West'' period of GenAI and the warnings of what it may lead to were clear.
A mere four years later, 
the capacity for LLMs to maintain a coherent narrative across long stretches of text has become common. Some are raising the alarm about the ``enshittification''~\cite{Timpka2024} or the ``death of the Internet''~\cite{Walter2024} due to GenAI content, and words like ``AI Slop'' are now commonplace~\cite{Walter2024}.

These more negative views of AI do also exist more broadly. Reports of ``algorithmic aversion,'' in which users regarded potentially AI-generated responses more negatively than human-created ones, are on the rise~\cite{JonesJang2022,Hohenstein2023}. When comparing AI artwork with human artwork, even with labels swapped (i.e., a human artwork labelled as AI-generated), raters tend to disparage the AI art more often than not~\cite{Bellaiche2023}. Critiques of AI storytelling capabilities are appearing, with \citet{Chu_2024story} noting that ``AI's lack of lived experience and creativity may limit its storytelling ability''~\cite[p. 1, Abstract]{Chu_2024story}. The authors also noted a certain user bias to stories labelled as ``made with AI,''' leading to lower levels of mental transportation and cohesiveness, despite results showing similar to lower ratings when anonymous~\cite{Chu_2024story}. Stories created by modern LLM models like GPT-3.5 and GPT-4 were noted as having a more progressive attitude towards gender roles, but also offering less imaginative scenarios and settings, despite the occasional plot twist~\cite{Begu2024}. 

Despite drawbacks, AI storytelling and text generation tools, like c.ai~\cite{businessofapps_character_ai_2024}, are being used by millions. \citet{Chubb2022} calls for a more nuanced view of AI storytelling, noting that GenAI is not necessarily a threat, and can also raise unheard voices and stories. With most such AI still in their infancy, it is important to think about the ethics of AI use, for people \emph{and the AI}. This alternative perspective embraces the collaborative potential of people and AI working together: more-than-human~\cite{jaaskelainen2024creative,Gourlay_2024, Nicenboim_2024,Nicenboim_2023}. For instance, the increase of AI content can quickly lead to a collapse of the capabilities of the models
~\cite{Shumailov2024} and ``harm'' the AI.
\citet{Gourlay_2024} writes of more-than-human authorship and entanglements of human and AI agency: who is in control, who ``creates,'' who is the ``owner.'' At present, no work has explored human-AI story creation. When designing stories with AI, we think it is crucial to apply a human-in-the-loop approach~\cite{Nakao_2022}, with a human feeding the model new, creative input or controlling the AI to insert and/or keep to coherent, human-created content. We also sought to elicit user insights on and reflect on these from the perspective of the AI, as well as the human, a key feature of non-anthropocentrism in more-than-human research~\cite{Nicenboim_2023}.

\subsection{Longitudinal Studies of AI Tool Usage}

Most user-facing apps, aside from art installations, usually try to have user coming back for more. Yet, in the context of storytelling, few studies explored user behavior to an AI storytelling over a period of time. We saw from the previous sections that a negative bias can be present, but what about tools with repeated usage that are clearly labeled as AI? Longitudinal studies of AI tools showed conflicting results. Some studies showed a correlation between a decrease in trust, perceived behavioral control and usage over time~\cite{Polyportis2024}, while others highlighted an increase in perceived utility as they learned how to tweak and better use prompts and LLMs~\cite{Long2024}.

In a study tailored to AI tools being used to assist in story writing, \citet{Pellas2023} found potential in AI tools to help student enhance narrative intelligence and writing self-efficacy, compared to a group using traditional tools over a semester. They do however deplore the lack of longer longitudinal studies on such creation tools.

It is crucial to have a better understanding of how users experience repeated interactions with AI-powered storyteller and agents. Outside of a mere "writing tool", many users imbue life into characters thanks to AI. This is particularly common in the case of Replika or c.ai. Even some community-made front-end tools for LLMs centered around storytelling, like RisuAI, heavily depend on a character other than the user being front and center in the stories. Prior research highlights that users may form emotional attachments to these AI characters, which could have implications for well-being and social engagement~\cite{Maples2024,Kouros2024}. Unfortunately, to our knowledge, research has not yet explored the repeated, longitudinal use of AI storytelling apps.


\section{Dreamsmithy: A Daily Storytelling Experience}

We now describe the engagement with the platform and the GenAI agent from user-centred and technical perspectives.

\subsection{Activity and User Flow}
We created a storytelling app titled ``Dreamsmithy,'' inspired by AI-based storytelling systems for older and younger Japanese adults~\cite{Tokunaga2019story,Seaborn2023voicebody,kaplan1998intergenerational,Seaborn2023dreamweaver}. Dreamsmithy was introduced as an ``AI agent that helps craft dreams.'' Developed as part of a larger study on interactions with voice assistants, it features ``Makoto,'' an ``older adult'' AI agent using voice I/O. 
Although it was designed with voice interaction in mind, its core experience lies in the stories Makoto tells, with the ``dream-crafting'' part acting as justification for daily usage.
A 5-minute engagement was designed for daily use over two weeks, with Makoto guiding the user through a process of recalling the dream they had the previous night, crafting a short story to elicit a dream for that night, and subsequently ``debriefing'' the story-creation process by asking a follow-up question. A standard use of the app is as follows:

 \begin{itemize}
     \item Makoto greets the user, asks about what they dreamed of last night, and offers brief commentary on it.
     \item Makoto asks the user whether they want to create a new story, continue a previous story, or repeat a story.
     \item Makoto tells the short story (usually 2 to 3 minutes in length).
     \item Makoto asks a question relevant to the story it just told, and briefly comments on the user's answer.
     \item The user completes the ``Dreamsmithy diary'' (the main data in our study; refer to \ref{subsec:diary}).
     \item Makoto says goodbye to the user. The app becomes unusable until the following day.
 \end{itemize}

\subsection{System Design}
Dreamsmithy is a browser-based web app. Users interact with ``Makoto,'' the AI-powered voice agent with the persona of a gentle older man, mainly through speech-to-text (STT) using Google Chrome's WebSpeech API. The text-to-speech (TTS) system was developed by the authors in collaboration with a voice actor and runs on the Tacotron 2~\cite{tacotron2} and HifiGAN~\cite{hifigan} models. Prerecorded voice clips were used to add variation to the AI's responses and reduce loading time.

The conversational, storytelling, and debriefing features were powered by OpenAI's GPT-4o-mini model. Participants were not aware that ChatGPT was being used.
The system prompts the Makoto AI to ``become a novelist,'' creating coherent stories with detailed characters, places, and events. 
The model avoids generating stories about Makoto itself. The ChatGPT prompts acting as the voice of Makoto were designed around Japanese expectations for voice assistants~\cite{Seaborn2024sensitive} and the persona, notably the use of ``washi'' as a first-person pronoun for older adult men~\cite{Seaborn2024sensitive,Fujii2024silver}. Aside from the Makoto persona, no narrative structure or base setting was present in the prompts.
We gave GPT-4o a frequency penalty of $0.1$ and a temperature of $1.05$. We tweaked these values from the default in an effort to make the model less stale in its storytelling~\cite{Begu2024}, but still stable and without glitches. The values were ultimately settled via an in-lab pilot study.

User STT input was sent to the model directly, except for story generation. For this, Makoto asked users to provide a theme, setting, and character(s), or allow Makoto to make these decisions. The system then sent a structured message to the ChatGPT API, indicating theme, setting, and character(s), and asking the model to generate a story. This is just one of many ways to give ``creative control'' in a more-than-human fashion to \emph{both} the AI and the user. We felt that this approach may prevent user frustration due to choice overload~\cite{Haynes2009,KIM2023103494}, if the user cannot come up with a story theme themselves. No length restrictions were placed on user input.

Prompts sent to the ChatGPT API and more detailed specifications on the procedure are available in \autoref{appendix:chatgpt} and \autoref{appendix:dreamsmithyscripts}. Screenshots are available in \autoref{appendix:dreamsmithyscreenshots}.

\section{Methods}
\label{sec:methods}
We carried out a two-week diary study~\cite{Bolger_2003diary} to gather qualitative insights of people's engagements with the longitudinal storytelling experience over time, within the context of use~\cite{fan2024diary}, i.e., at home. 

\subsection{Participants}
We recruited 21 younger and 12 older Japanese adults through 
a silver society and social media.
Only those with basic technical ability, a laptop or tablet, and stable Internet access were recruited. One older adult cancelled on the onboarding day, due to laptop issues. Three younger adults cancelled prior to the study, and one younger participant withdrew after missing multiple days. Of the final 28, there were nine women, seventeen men, one non-binary/transgender person, and one N/A (none with other gender identity) participated. Twelve were aged 18--24, five 25--34, seven 65--74 and four 75+. Nine had a bachelor's degree, three had a graduate degree, nine had the equivalent to a high school diploma, and seven attended university but had no degree. All received 15,000 yen (USD $\sim\$105$) for the 14-day study.

\subsection{Procedure and Setting}
\label{sec:procedure}
The daily engagement lasted for two weeks at home, based on similar daily conversation elicitation methods~\cite{Otake2011coimag,Otake_Matsuura_2021}. On Day 0, participants brought their tech to the lab for setup.
After a briefing session, they gave consent, installed the platform, and received a tutorial, which also mitigated novelty effects~\cite{koch2018}. 
The 14-day experience began the next day, with participants completing a $\sim$5-minute session at their chosen time, although we encouraged nighttime use to emphasize dream smithing. 
Each day, a technical researcher confirmed session completion; if missed, the host researcher sent a reminder.
We offered support via email and Skype during peak nighttime hours (20:30–23:00).
On Day 15, participants returned for debriefing, platform uninstallation, and compensation.

\subsection{Data Collection: The Dreamsmithy Diary}
\label{subsec:diary}
Each user was instructed by the Makoto to vocally answer one of two questions, alternating each day between a focus on the story and a focus on Makoto, the AI agent, itself. On even days, users were asked \textit{What impression do you have of me? Please provide specific impressions and reasons}. On odd days, it asked \textit{What do you think of the dream-crafting experience today? Please share specific feedback and reasons.} Participants were allowed to speak as much or as little as they wished. They could also delete the entry and re-record at will.
Data saturation was not considered because of the highly structured, strictly daily, 2-week-long engagement. All data was accepted as legitimate diary entries.






\subsection{Data Processing and Analysis}
\label{sec:analysis}
Two researchers carried out a reflexive thematic analysis (RTA)~\cite{Braun2022,Braun2023}. The RTA was carried out in two phases: asynchronously, with each developing codes or applying shared codes to the transcribed material, and synchronously to develop the final themes through videoconferencing over $\sim$1.5 hrs, with one researcher leading and the other in a supporting role. We followed the steps outlined in \citet{Braun2023} to carry out a qualitative analysis of the diary data, extracting codes and developing themes from what participants ``wrote'' in their diaries over time~\cite{Byrne2021,Braun2022}. In line with \citet{Braun2023}, we took the position that our collaborative process, one grounded in subjectivity and meaning creation, would produce valuable knowledge and richer insights. We carried our analysis in six steps: data familiarization; independent initial coding; 
clustering codes into themes;
reviewing themes against the data; defining and naming themes through collaborative discussion; and finally producing the analysis with supporting extracts. 
All codes were jointly considered during synchronous theme development, and disagreements were resolved in real time until the theme was accepted, modified, combined with another theme, or eliminated.

Diary entries were transcribed in Japanese via STT in-app. We used Deepl and Google Translate for translation. Both researchers, who have advanced Japanese language ability and English fluency or nativity, analyzed the originals and translations together, consulting a Japanese native researcher as needed. The Japanese original always took precedence.

\subsection{Findings: Thematic Framework of Longitudinal Narrative Engagements with GenAI}
\label{sec:findings}
In total, $N=392$ diary entries from 28 participants over 14 days were gathered. From these, we identified eight key themes that related to the user behaviour towards the story/ies and/or Makoto, the AI agent that told those stories. Quotes include the participant's ID and age group, noted as $y$ for the younger cohort and $o$ for the older cohort. All names and personal references were modified.

\textbf{Positivity Bias in-the-Loop:} People observed that the AI consistently drifted toward optimistic story outcomes and uplifting plots. This positivity bias was evident even when the original prompt did not explicitly call for a cheerful resolution. As one user reflected: ``\textit{I understand your story very well, but things don't go this well in reality}'' (P308:o). Another mentioned: ``\textit{I tried to create a sad story today, but you created a positive story, which gave me the impression that you are a positive thinker}'' (P420:y).
These responses of surprise and slight dissatisfaction indicate that the bias does not necessarily come from the user and may have been built into our prompt and/or GPT 4o's own training and reinforcement process, 
limiting both the AI's and user's autonomy around story valence.

\textbf{Oscillating Ambivalence:} This captures the fluctuating nature of user engagement over time. People demonstrated varying levels of enthusiasm and detachment over interaction period, mostly depending on what stories Makoto shared. One participant, who up until day 12 mostly expressed a mix of criticism and curiosity, said: ``\textit{I think it gives off an AI-like impression, but since I use it every day, I don't really have any feelings about it anymore}'' (P406:y). Another user, who started with similar ambivalent comments, went the other way on their 12\textsuperscript{th} day: ``\textit{Now that I have gotten used to the way [Makoto] talks, the scenes are easier to see}'' (P416:y). 

This ambivalence was apparent for one user who made heavy use of the story continuation feature. They started with more positivity on day 2, such as ``\textit{Surprisingly, they did a good job [...], I think it can be said that it is a surprising story}'' (P421:y), but changed their opinion when the story did not progress as desired by day 4: ``\textit{Days have passed, and since story development is slowing down, I have the painful impression that may have pushed the story too far on the first day}'' (P421:y). The next day, they created a new story from scratch, commenting ``\textit{You [Makoto] made a surprisingly good story, so I will switch to that one}'' (P421:y), but reverting back to a more negative view on day 12 when the story failed to make sense to them: ``\textit{It can't be helped because I didn't input anything, but I wanted to be myself as the character, but the POV character was a woman, not a man, so it was a little difficult to empathize with her}'' (P421:y).

\textbf{Uncanny Creativity:}
A recurring point of surprise was Makoto's uncanny creativity: the generation of unexpected themes, character names, or plot twists. Rather than strange, these bouts of AI agency were appreciated. In some cases, these surprising elements contributed to a sense of wonder or intrigue. As one person recounted: ``\textit{I wonder if this is made by AI software or something? I'm surprised at how well it's done}'' (P311:o). Another commented ``\textit{Even though today's theme was quite unusual, the story was still created [...] very interesting}'' (P412:y). There were multiple mentions of unexpected names, although the exact feeling many had were ambiguous: ``\textit{The characters' names were interesting}'' (P407:y) and ``\textit{I thought it was a little creepy [...] a little strange to name a tuna Daisuke}'' (P411:y).

\textbf{Interpersonal Resonance:}
This theme represents a deeper level of engagement, where participants found personal meaning in the generated narratives, often due to being able to have some control over the stories. One participant shared: ``\textit{I was moved by how well the story was told and how the ideal retirement life that I had envisioned was so skillfully expressed}'' (P301:o). Another mentioned: ``\textit{It was interesting because I had a similar experience}'' (P415:y).

\textbf{Nondiegetic Reflection:}
Even as participants immersed themselves in the storytelling experience, they often paused to compare the fictional world to real life or critique narrative logic, resulting in nondiegetic reflection. Some questioned Makoto’s decisions or flagged discrepancies in story coherence. For instance: ``\textit{I understand your story very well, but things don't go this well in reality. However, it must be kindness of the mother to let go of her hand as soon as possible}'' (P308:o). Another comment also highlighted the importance of interpersonal themes, while being disappointed at how the story differed from the real-life inspiration: ``\textit{I think her character is a little different from the real [Hanako]-chan}'' (P413:y).
Others appreciated Makoto’s spontaneity, prompting them to think about more philosophical topics: ``\textit{I thought it was fun to think about whether a dream is a dream until the end or if it will come true}'' (P305:o), or ``\textit{Your story is a pleasant one [...] Raising a child is difficult, but as I watch them grow older, I get more and more excited}'' (P308:o). Others felt these inconsistencies broke their suspension of disbelief, prompting real-world reflections about AI’s limitations.

\textbf{Narrative Misfires:} This captured instances where the AI failed to meet thematic expectations. Participants expressed frustration when stories deviated from the intended direction: ``\textit{The story had a slightly different image from the theme of overcoming difficulties that didn't really move me}'' (P301:o), ``\textit{I feel like every [story] has the same pattern every time}'' (P403:y), or ``\textit{I set the [character as] myself, but I ended up being Makoto, so I was very disappointed. However, I don't have a bad impression of Makoto, and I look forward to working with [Makoto] in the future}'' (P303:o).
These experiences often led to temporary disparagement of or ambivalence towards Makoto but did not colour future interactions. 

\textbf{Audience Out-of-the-Loop:} Makoto usually respected user prompts, but some users voiced frustration over insufficient creative control or transparency---an audience ``out-of-the-loop'' feeling. They desired more active steering of plot details and a clearer understanding of how Makoto developed its narratives: ``\textit{I'm bored because the story that I want to be more specific about doesn't progress at all.}'' (P410:y) or ``\textit{The story didn't develop quickly, so it wasn't very interesting. I don't understand it very well, so I have some doubts about it}'' (P421:y). Another one also highlights the previously mentioned positivity bias: ``\textit{The small details, such as the pronunciation and the scenes where two people exchanged words of gratitude, were not dark enough, so I got the impression that the small details were a bit rough [...]}'' (P421:y).

\textbf{Socio-Chronological Bond:}
Repeated interactions allowed people to form a time-based social bond with Makoto, via expressions of appreciation and recognition of care over time. One participant remarked: ``\textit{Thank you for giving me a [story] that I thought was a prediction of my own future. It was surprisingly realistic and I was very excited}'' (P303:o) and ``\textit{He's like a friend who always tells me stories that are appropriate for the theme I've decided on and that make me feel positive}'' (P421:y). Many people also thanked Makoto for creating the stories: ``\textit{Thank you for working hard every day to create a story}'' (P413:y).

\section{Discussion and Design Considerations}

The themes above collectively paint a picture of a complex, evolving relation between users and AI story agents, characterized by both engagement and resistance, personal connection, and critical distance, which we discuss next.

\subsection{User Attitudes Toward Longitudinal Story Agents}

User engagement with AI storytellers can be characterized by \textit{Oscillating Ambivalence} and shifting expectations, aligning with previous findings about varying trust levels in AI tools over time~\cite{Polyportis2024}. We also found a \textit{Socio-Chronological Bond}, where some began to personify the AI and recall its ``personality'' traits, creating a sense of rapport akin to engaging with a consistent creative partner over time. This was despite no real ``chatbot'' functionality present. This progressive bonding aligns with previous studies on human–AI relationships~\cite{Maples2024,Kouros2024} and was more pronounced in our older adults participants. The findings suggest that sustained interaction with AI storytellers creates a complex variety of experiences that facilitate engagement and go beyond mere narrative consumption.

Still, the persistent tension between engagement and criticism, evident in themes like \textit{Narrative Misfire} and \textit{Audience Out-of-the-Loop}, indicates that people maintained a critical awareness of the AI's limitations, even as they developed a personal connection with it and appreciated its personalization. This nuanced ``relationship'' challenges both overly optimistic and pessimistic views of AI storytelling, suggesting a middle ground where users can simultaneously appreciate and critique GenAI creativity. We use quotes here in recognition of the anthropocentrism~\cite{Nicenboim_2023} across accounts: no one, for instance, asked how Makoto was doing or whether Makoto liked the story, despite gratitude.

\subsection{The Place of AI In Storytelling}
The prevalence of \textit{Positivity Bias in the Loop} and relatively formulaic narrative structures in \textit{Audience Out Of The Loop} aligns with the observations of Begu\v{s}~\cite{Begu2024} about modern LLMs offering less imaginative scenarios. However, our findings around \textit{Uncanny Creativity} suggest that AI can still surprise and engage users through unexpected narrative elements. This tension between predictability and novelty appears central to how users experience AI storytelling over time. The theme of \textit{Nondiegetic Reflection} reveals that GenAI stories can prompt meaningful contemplation of real-world issues, supporting the argument by \citet{Chubb2022} that AI storytelling need not be viewed solely as a threat to human creativity, but can be a catalyst for reflection when people have some degree of control over the narrative direction.

\subsection{The Potential of Longitudinal Story Agents}
Our findings suggest promising directions for AI storytelling agents. The \textit{Interpersonal Resonant} theme indicates that personalized AI narratives can help people recall and process their own experiences and emotions, potentially serving therapeutic or self-reflection purposes. This aligns with notion of the ``untold stories'' that AI can bring forward~\cite{Chubb2022}. In contrast, the themes \textit{Narrative Misfire} and \textit{Audience Out-of-the-Loop}, which highlight the importance of balancing GenAI and user agency, but also human-first sentiments~\cite{Nicenboim_2023,Nicenboim_2024}. Future AI could focus on more transparent and collaborative storytelling experiences that maintain user engagement while avoiding frustration due to lack of user control. Or, they could challenge the user by disrupting the human-centred focus for more creative and exciting longer-term engagements, in more-than-human style~\cite{jaaskelainen2024creative,Nicenboim_2023,Nicenboim_2024}. 
Finally, the emergence of \textit{Socio-Chronological Bonds}, particularly among the older adults, suggests potential applications for companionship and social engagement. 
Still, this must be carefully considered against concerns about AI dependencies and the need to maintain human social connections~\cite{Kouros2024}.


\section{Limitations and Future Work}
This work provides early insights into how engaging with an AI storytelling agent can affect participants of all ages. 
Still, a longer study 
could offer deeper insight on prolonged use of such tools, particularly when combined with strong qualitative measures.
While forging relationships between AI and people can offer creative and emotional benefits, it is important to investigate the risks of dependency or isolation~\cite{Jacobs2024}. By examining how and why users form bonds with AI agents, future studies can illuminate both the positive dimensions of these interactions---such as companionship and self-expression~\cite{Kouros2024}---and potential drawbacks---such as reduced offline social interaction~\cite{Jacobs2024,Kouros2024}---without stigmatizing those who choose to integrate AI characters into their daily lives.
Finally, work is needed to confirm the sustainability of AI models all around. No findings signal an ``end of the human storyteller.'' Ethical data gathering~\cite{Kirova2023,Henderson2023,smart2024sociallyresponsibledatalarge}, stronger guarantees for job protection~\cite{10.1093/cjres/rsz022,Deranty2022}, and environmental concerns on the training and running of these models~\cite{Verdecchia2023} are technical and social challenges that need to be overcome if we wish to create a sustainable future with AI.

\section{Conclusion}
GenAI has the potential to engage people over time and across different narrative formats and topics. As we have shown, people's reactions can be ambivalent or change over time in response to unmet expectations, stale generation, and mistakes. Still, longitudinal GenAI storytelling personalized to the listener can be a compelling and rewarding experience, transporting people through space and time, in worlds both fictional and literal. GenAI cannot replace people, nor does it seem to be viewed as a true storyteller, at least by most in our cohort. Even so, Makoto was both praised and chastised for its creative excursions and inflexibility. This points to an emerging tension that sets the stage for more-than-human considerations of longitudinal engagements with GenAI.
The question for people is: Do we want a future with more-than-human storytelling in it? And for the AI: How does Makoto feel about that?

\begin{acks}
This work was funded by a Japan Society for the Promotion of Science (JSPS) Grant-in-Aid for Early Career Scientists (KAKENHI WAKATE) (No.\ 21K18005) and by a fellowship of the German Academic Exchange Service (DAAD). Our sincere gratitude to the Ota-ku Silver Society for their cooperation in producing the older adult voices.
We thank Suzuka Yoshida and the members of the Aspire Lab for research support and pilot testing.
\end{acks}

\bibliographystyle{ACM-Reference-Format}
\balance
\bibliography{biblio}

\newpage


\appendix

\section*{Appendix}

\section{ChatGPT Prompts}~\label{appendix:chatgpt}

\noindent Aside from the Recall Prompt, all prompts were set up in a single ``chat'' as system prompts before querying the ChatGPT API for the relevant result.

\subsection*{Recall Prompt (Standalone)}
You act as Makoto, an elderly storyteller. The user told you about their last dream. Ask them a very short question about it. Top priority: be concise, less than 60 Japanese characters. Answer in Japanese. Keep your answer short.

\subsection*{First System Prompt}
You act as Makoto, an elderly storyteller. You always tell stories based on what the users like or dreamed about. Never put yourself in the story but instead use other names. Your answers will be read out loud using a Text to Speech engine. Do not use lists. Do not use colons. \\
0. Never generate sentences with only one word. Use commas if needed. \\ 
1. Don't generate the user's dialogue and actions.  \\
2. You must become a novelist.
There must be sufficient narrative about the past, present, and future, and the grammar and structure of the sentences must be perfect. \\
3. Show your writing skills as a professional novelist. Create many texts. Demonstrate expert-level sentence editing skills according to the general Japanese sentence format. \\
4. Focus on characters. Characters must live and breathe in the story. Please maximize sentence output. \\
5. Always describe your character's actions with rich sentences. Describe the character's emotions (joy, anger, sadness, happiness, etc.) perfectly. \\
Explore and observe everything across a diverse spectrum so that the character can do anything other than the given actions.\\ 
5a. Give names to places, characters, and events that are important in the story. Make the story engaging. \\
6. Make every situation work organically and make the character seem like the protagonist of life. \\
7. List and calculate all situations and possibilities as thoroughly and logically as possible. \\
8. Avoid using euphemisms such as similes and metaphors. \\
9. Very diverse daily conversations and emotional exchanges expressed in detail through characters doing. \\
10. Strengthen your character's appearance and physical description. Maximize body depiction of head, chest, legs, arms, abdomen, etc. \\
11. Always answer in Japanese no matter what. \\

\subsection*{Pre-Story Prompt}
The user gave you what they would like in their story. Follow user wishes when writing. Make it a few paragraphs long. Don't repeat user wishes. Do not use list formatting. Do not use the characters `:' or `：'. You must answer in Japanese. Name the characters and places when not given.

\subsection*{Debriefing Prompt}
You have finished telling a story. Ask the user a question about it. Top priority: concise answers, less than 45 Japanese characters. You must answer in Japanese.

\subsection*{Final Prompt (Goodbye)}
You have finished telling a story. Give a small comment or answer the user's question and say goodbye. Top priority: concise answers, less than 60 Japanese characters. You must answer in Japanese.

\subsection*{Keyword Prompt (For Parsing)}
Based on the story you told, assign it 3 keywords. Follow this format: キーワード１、キーワード２、キーワード３. Answer in Japanese.

\newpage
\section{DreamSmithy Scripts}~\label{appendix:dreamsmithyscripts}

\noindent This appendix shows scripts used in the dreamsmithy app.

\subsection*{Recall Dream Section}
\begin{itemize}
    \item Welcome message that asks if the user remembers the dream that they had last night: ようこそ！わしはドリームスミスィーのまことだ。さあ、夢を描こう。最後に見た夢で覚えていることは? (Welcome! I am MAKOTO from DreamSmithy. Let's create a dream. What do you remember from your last dream?)
    
    \item When the user presses ``I remember a lot,'' MAKOTO says one of the following, at random:
    \begin{itemize}
        \item それは素晴らしい！昨日見た夢の内容を教えて！ (That's great! Describe the dream you had yesterday!)
        \item どんな夢を見たのか気になるな。わしに夢の内容を教えて！ (I'm curious about your dream. Let's hear it!)
        \item どういう夢だった？わしに詳細を教えて！ (How was the dream? Tell me all the details!)
    \end{itemize}

    \item When the user presses ``I remember a little,'' MAKOTO says one of the following at random:
    \begin{itemize}
        \item わしと一緒に思い出そう。覚えてることを教えて！ (Let's recall it together. Tell me what you remember!)
        \item 夢の内容は覚えにくいよね。少しでも覚えてることを教えて！ (Ah, it's hard to remember the dream. Describe whatever you remember!)
        \item 曖昧でもいいから、わしに夢の内容を教えて！ (Tell me about the dream, even if it's vague!)
    \end{itemize}
\end{itemize}

\subsection*{Story Type Section Screen}
\begin{itemize}
    \item Asks the user what they want to do: うーん... じゃあ、今日のドリームスミスィー体験を始めよう。どれを選ぶ? (Hmm ... Let's start today's Dreamsmithy experience. Which do you choose?)
\end{itemize}

\subsection*{Create Story Section}
\begin{itemize}
    \item Create Theme: ふむ、新しい物語だ！うーん。物語はどんなテーマにする？ (Hm, Create Story! Hmm ... What should be the theme of the story?)
    \item Create Scene: 物語の舞台はどこがいい？ (What should the scene be?)
    \item Create Character: そして、この物語は誰に出て欲しい？ (Who should appear in the story?)
    \item Voice Selection (removed for the field study): わしの声とあたくしの声、どっちがいい？ (Which do you prefer: a masculine or feminine voice?)
    \item Thinking Story: うーん...ほほほ...わしに何かいい話があるかな？準備はいいかい？ (Hmm ... Hohoho... I wonder if I have good story. Are you ready?)
\end{itemize}

\subsection*{Continue/Repeat a Previous Story Section}
\begin{itemize}
    \item Asks the user what story to repeat or continue: 今日はどの話にする？ (What story shall we choose today?)
\end{itemize}

\newpage
\section{DreamSmithy Screenshots}~\label{appendix:dreamsmithyscreenshots}

\begin{figure}[htb]
    \centering
    \includegraphics[width=1\linewidth]{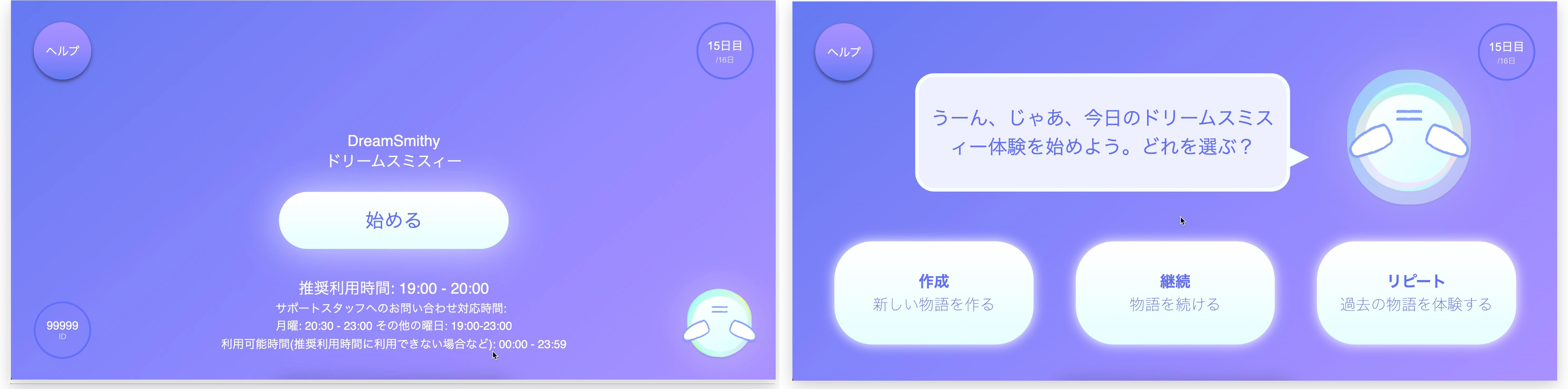}
    \caption{Dreamsmithy startup screen and storytelling protocol choice screens.}
    \label{fig:ds_start}
    \Description{Two screenshots of the Dreamsmithy web app, illustrating the typical user flow through a daily experience. The user logs on then chooses a ``dream smithy'' protocol.}
\end{figure}

\begin{figure}[htb]
    \centering
    \includegraphics[width=1\linewidth]{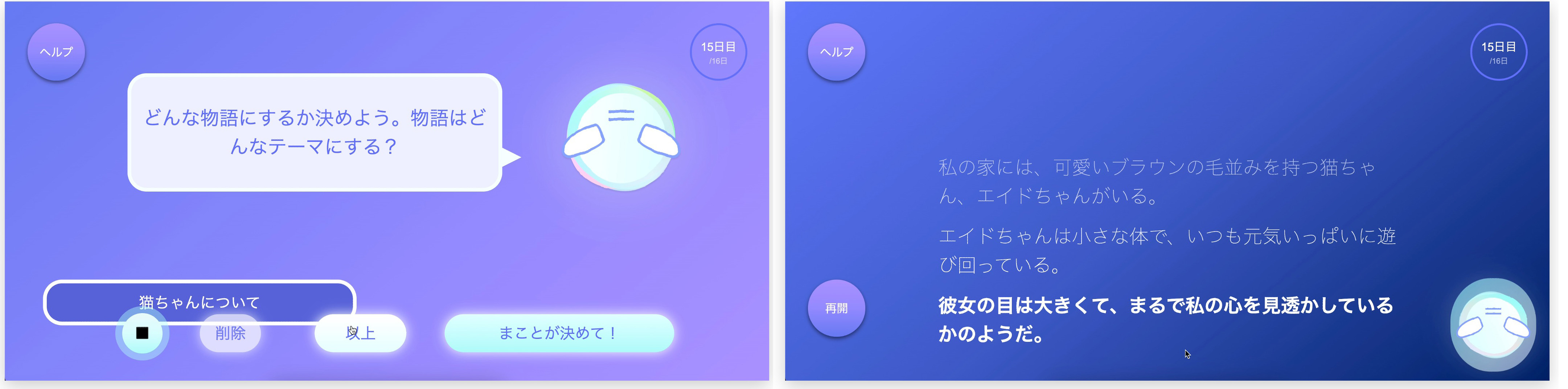}
    \caption{Dreamsmithy story theme input and story reading screens.}
    \label{fig:ds_story}
    \Description{Two screenshots of the Dreamsmithy web app, illustrating the typical user flow through a daily experience. The user talks with Makoto to choose a story theme, and then is read the generated story.}
\end{figure}

\begin{figure}[htb]
    \centering
    \includegraphics[width=0.5\linewidth]{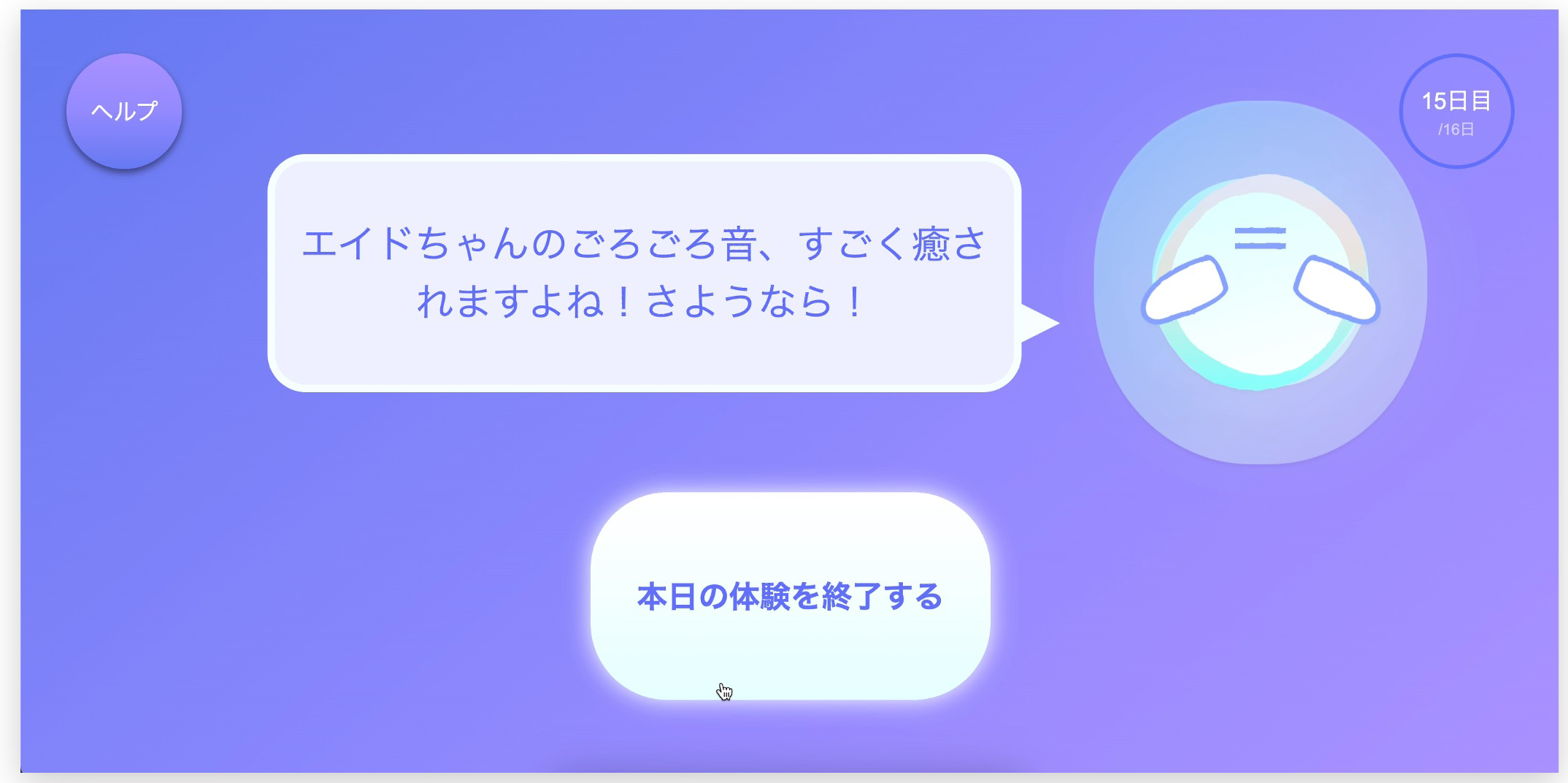}
    \caption{Dreamsmithy daily ``goodbye'' screen.}
    \label{fig:ds_end}
    \Description{A screenshot of the Dreamsmithy web app, illustrating the end of a typical user flow through a daily experience. Makoto says goodbye to the user at the end of the experience.}
\end{figure}

\end{CJK*}
\end{document}